\documentclass[prb,preprint,superscriptaddress,amsmath,amssymb]{revtex4}

\usepackage{graphicx}%
\usepackage{dcolumn}%

\begin{document}

\title{Angle-resolved photoemission using the circularly polarized light in Pb-Bi2212.}

\author{S. V. Borisenko}
\affiliation{Institute for Solid State Research, IFW-Dresden, P.O.Box 270116, D-01171 Dresden, Germany}

\author{A. A. Kordyuk}
\affiliation{Institute for Solid State Research, IFW-Dresden, P.O.Box 270116, D-01171 Dresden, Germany}
\affiliation{Institute of Metal Physics of National Academy of Sciences of Ukraine, 03142 Kyiv, Ukraine}

\author{A. Koitzsch}
\author{M. Knupfer}
\author{J. Fink}
\affiliation{Institute for Solid State Research, IFW-Dresden, P.O.Box 270116, D-01171 Dresden, Germany}

\author{H. Berger}
\affiliation{Institut de Physique de la Mat\'erie Complex, Ecole Politechnique F\'ederale de Lausanne, CH-1015 Lausanne, Switzerland}

\date{\today}
\begin{abstract}
In a recent preprint Campuzano et al.\cite{CKV} have questioned the validity of our ARPES results\cite{CPOD} contradicting the interpretation of earlier photoemission experiments in terms of the time-reversal symmetry breaking in Bi2212\cite{Kaminski,Simon}. Here we highlight the principal results of our study and refute all the criticism.
\end{abstract}

\maketitle

To observe a time-reversal symmetry breaking caused by circulating currents in the pseudogap regime of cuprate superconductors an ARPES experiment using circularly polarized light was proposed\cite{Varma_old}. First experimental studies aiming to test this proposal gave negative results: the predicted effect\cite{First_dichro} was observed at various temperatures and in differently doped samples thus rulling out its clear relation to the pseudogap phase\cite{BorisDichro}. However, recent observation of a small ($\sim 4\%$) dichroism in the ($\pi$, 0)-point of the underdoped Bi2212 upon entering the pseudogap regime has been interpreted in terms of the time-reversal symmetry violation\cite{Kaminski}. For this purpose, a new pattern of circulating currents was introduced, now breaking the reflection symmetry in the $x=0$ and $y=0$ mirror planes\cite{Simon}, contrary to the previously predicted symmetry breaking in $x=\pm y$ mirror planes\cite{Varma_old}. In spite of several reformulations\cite{CKV,Kaminski,Simon}, the new criterion, suggested to judge whether the time-reversal symmetry is violated or not, is unequivocal: {\it if the dichroic signal measured in the pseudogap state of an underdoped cuprate and corresponding to the photoemission within the $x=0$ and $y=0$ mirror planes is finite, the time-reversal symmetry is broken \cite{Planes}}. It is this criterion which we have exploited in our study\cite{CPOD}.

Formally speaking, this implies that one can solve the problem by simply measuring the photoemission intensity in, e.g., the ($\pi$, 0)-point using the right- and left-hand circularly polarized light and then consider the difference (dichroism). We call this a "key experiment". Obviously, in such a key experiment the following conditions should be satisfied: (i) no geometric chirality\cite{Planes} in the experimental setup; (ii) the light must be circularly polarized; (iii) the sample must be in the pseudogap state; (iv) the momentum of the outgoing photoelectrons should lie exactly in the crystal mirror plane. We show below that our experimental conditions fulfil all requirements and our results clearly indicate the absence of the effect in underdoped Pb-Bi2212.

To facilitate the comparison with the previous work, we have chosen a similar experimental geometry (see, for instance, Fig.1 in Ref. \onlinecite{CPOD}). In such an arrangement one measures the dichroic signal in a number of {\bf k}-points introducing the geometric chirality of both signs and thus definitely fulfilling condition (i) for at least one direction. Condition (ii) is satisfied by utilizing the light from the wiggler-undulator beamline at ELETTRA which delivers $\sim 90 \%$ circularly polarized radiation with the photon energy of 50 eV. 

\begin{figure}[t!]
\includegraphics[width=7.9cm]{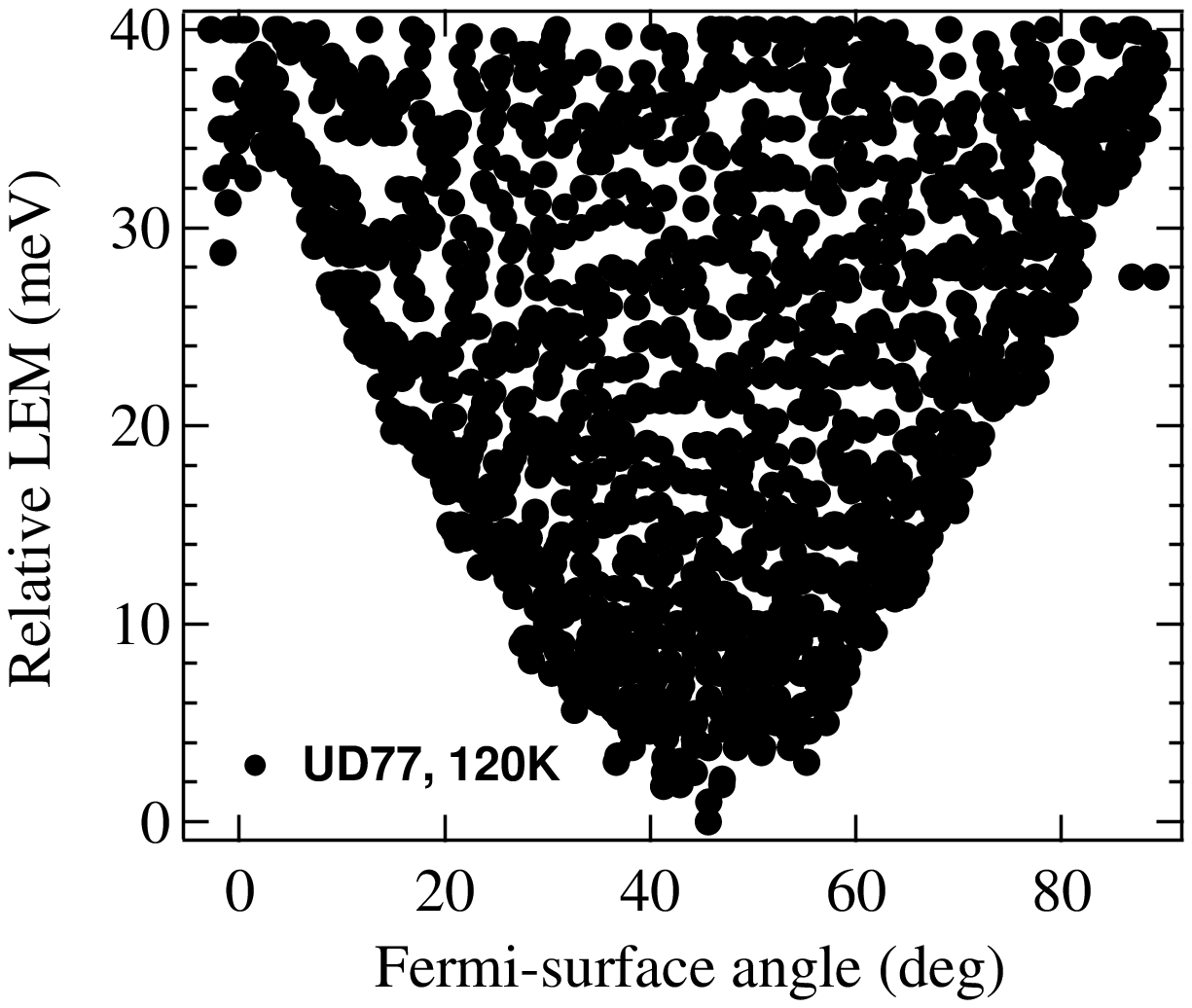}
\includegraphics[width=8cm]{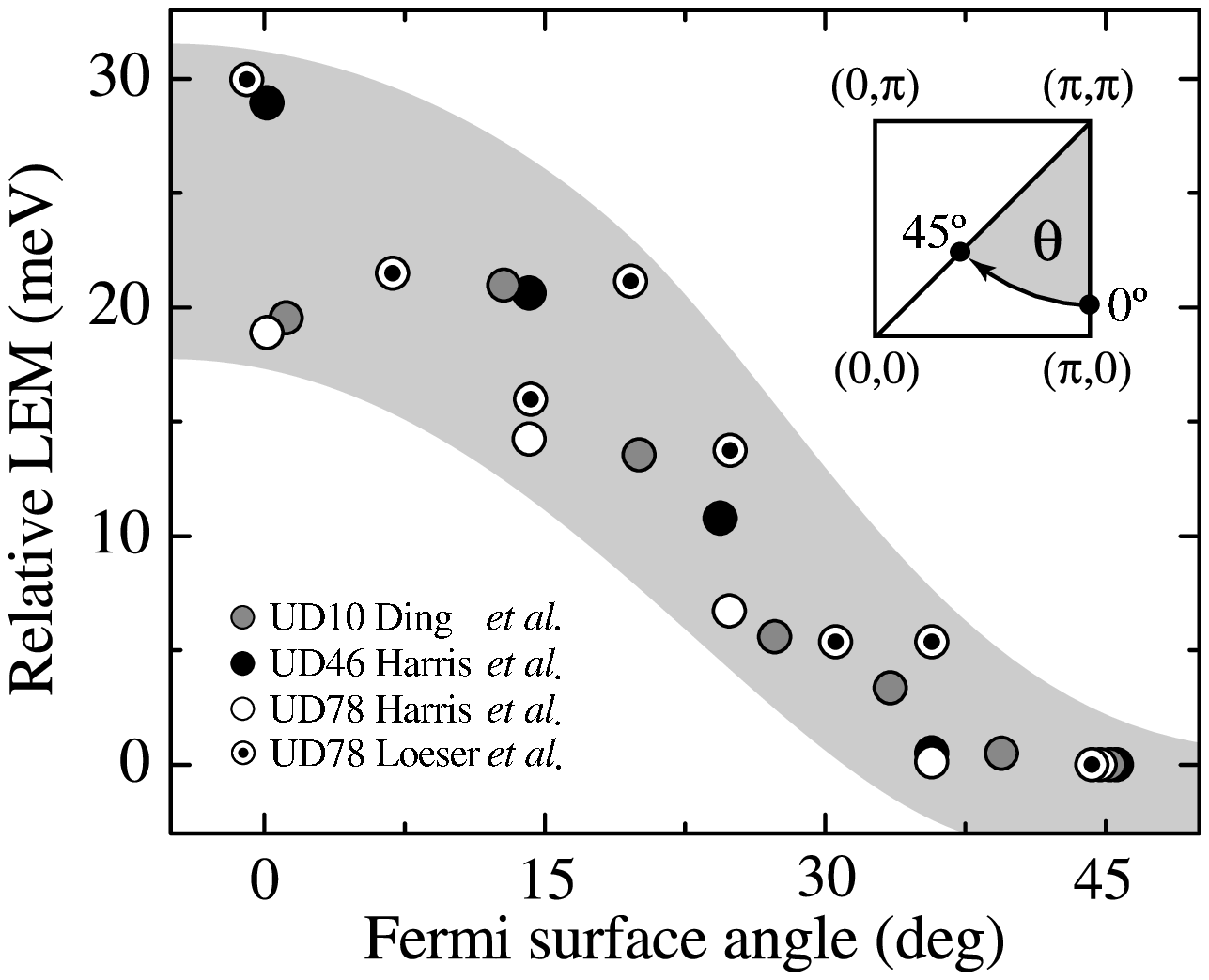}
\caption{\label{Pseudogap} Left panel - values of the EDC's leading edge midpoint (LEM) as a funtion of the Fermi surface angle within the quadrant of the Brillouin zone (see inset). The curve joining the low-gap extremity of these data points would represent the {\bf k}-dependence of the pseudogap. Right panel - collection of the similar data taken from Ref. \onlinecite{Damascelli}.}
\end{figure}

As stated in Ref. \onlinecite{CPOD} and clearly seen in Fig.~\ref{Pseudogap}, we have observed an anisotropic normal state pseudogap in our samples. Moreover, the size of the pseudogap and its momentum dependence are in a good agreement with the data published before\cite{Damascelli}.

In order to fulfil the condition (iv) it is crucially important to fix a momentum scale, since the value of the dichroism determined for the {\bf k}-points {\it known} to be in the mirror plane is decisive. In other words, that single direction for which condition (i) holds must be in the mirror plane. For this purpose we have recorded the spectra from a sufficiently large {\bf k}-space region which included clearly dispersing features, symmetrical about the mirror plane (for details see Ref. \onlinecite{CPOD}). This procedure of the mirror plane determination is a distinguishing feature of our experiment.

\begin{figure}[t!]
\includegraphics[width=13cm]{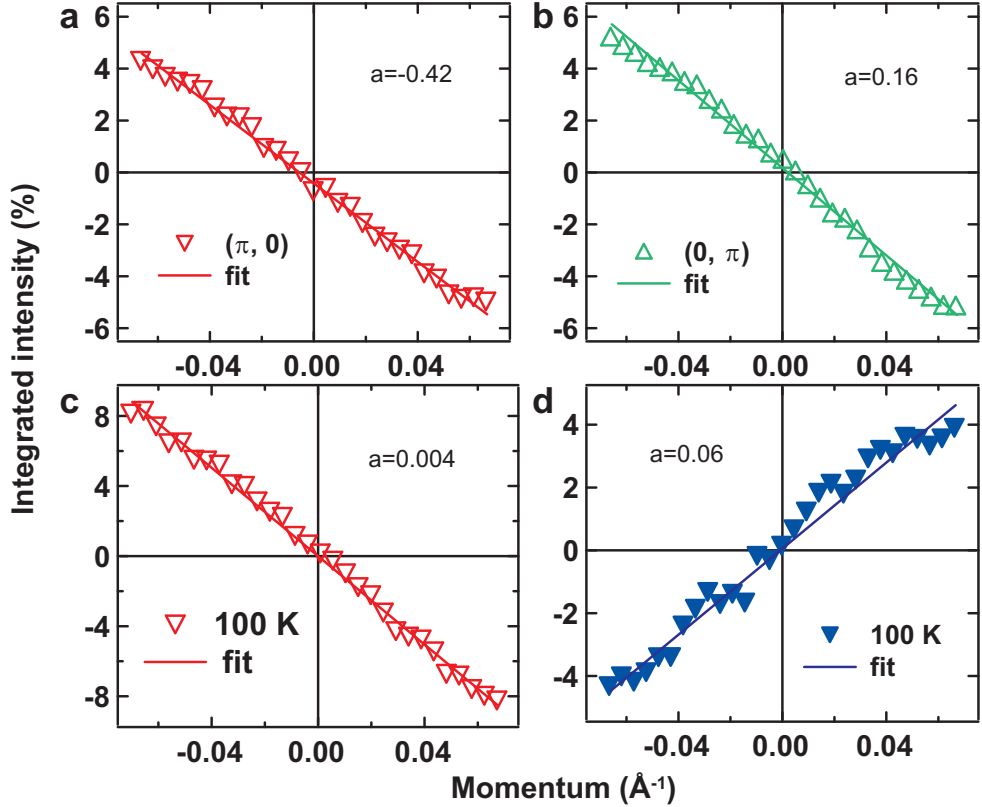}
\caption{\label{Key} Data presented in Ref. \onlinecite{CPOD} for the UD77K sample in the pseudogap regime together with the linear (a+bx) fits. a) ($\pi$, 0)-data from Fig. 3c, b) (0, $\pi$)-data from Fig. 3c, c) 100 K - data in the narrower momentum region from Fig. 3d, d) normal incidence data from Fig. 3f. The values of the fit parameter {\bf a} are also given.}
\end{figure}

\begin{figure}
\includegraphics[width=13cm]{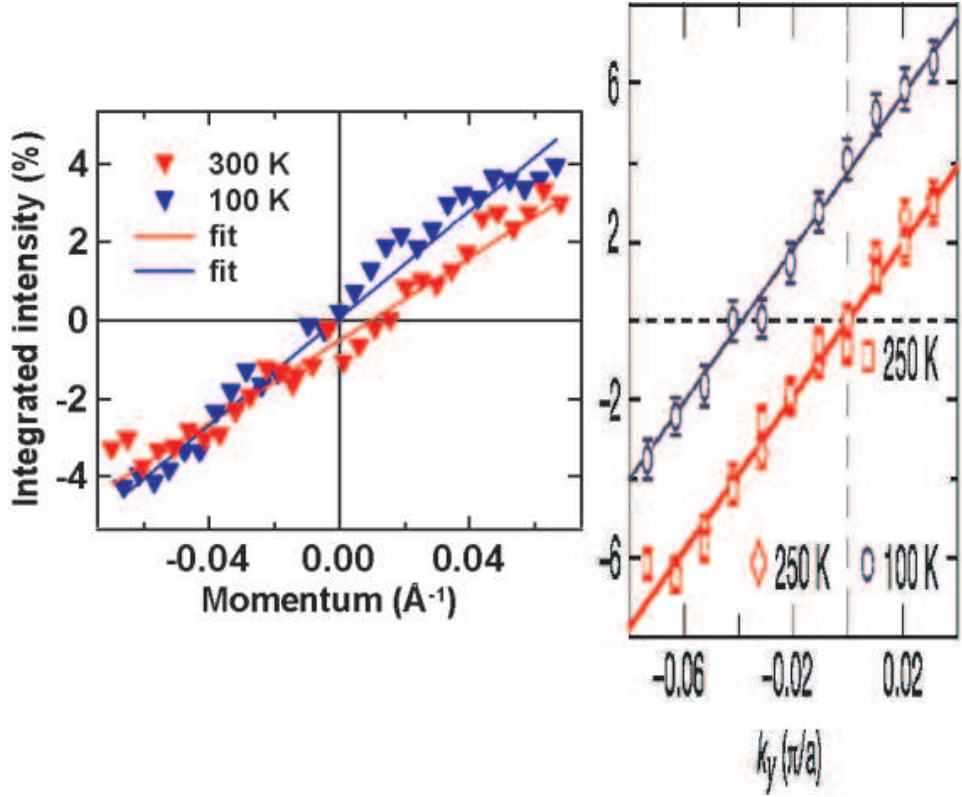}
\caption{\label{Comparison} Left panel: data from Fig. 3 e and f in Ref.\onlinecite{CPOD} together with the linear (a+bx) fits. Right panel: data from Fig. 3 g in Ref.\onlinecite{Kaminski}. Data points for the intermediate temperatures are removed for clarity. The plot is sqeezed to match the scales in the panels.}
\end{figure}

We have performed tens of the aforementioned key experiments. The results of four typical key experiments on UD77K samples at 100 K have been presented in our manuscript\cite{CPOD} and are shown in Fig.~\ref{Key} here together with the auxilary data from other {\bf k}-points. The size of the dichroism in the mirror plane is given by the value of the fit parameter {\bf a} (see caption to Fig.~\ref{Comparison}). These values are all within the estimated average accuracy of the experiment $\sim 0.5\%$. As easily seen from Fig.~\ref{Key}, one can safely conclude the absence of any systematic deviations from zero for the {\bf k}-points corresponding to the mirror planes.

We want to stress here once again that our results cannot be considered as manifestation of the time-reversal invariance of the electronic states in Pb-Bi2212. Using the clearly superior experimental setup compared to that of Kaminski et al.\cite{Kaminski} we observe a vanishingly small value of the dichroism in the mirror planes in our 5x1 superstructure-free underdoped Pb-Bi2212 samples, thus questioning the interpretation\cite{Simon} of earlier experiments in terms of the time reversal symmetry breaking.

Now we deal with the critical comments by Campuzano et al.\cite{CKV} (hereafter referred to as CKV).


The main concern of CKV is that no evidence was presented regarding the magnitude of the pseudogap in our samples. Once again showing in Fig.~\ref{Pseudogap} well known to the community results, we demonstrate that there can be no more doubts in this respect. We are not aware of the technique which CKV used to arrive at the conclusion that the pseudogap in our Pb-Bi2212 samples is "unnoticeable", comparing the room temperature (!) intensity map taken on the overdoped sample (Fig. 1 e in Ref. \onlinecite{CPOD}) with the 100 K intensity maps recorded some 1/2 of the BZ size away on the underdoped sample (Fig. 3 a,b in Ref. \onlinecite{CPOD}). In any case, considering the data presented in Fig.~\ref{Pseudogap}, this technique yields apparently misleading results.

The tendency to compare actually incompatible things is again clearly seen in the following examples. CKV make two remarks as for the error bars (page 4 in Ref. \onlinecite{CKV}) presumably intended to uncover an inconsistency between the given values. These values are indeed very different because the precision of 0.002 \AA$^{-1}$ stated in the manuscript (not 0.001 \AA$^{-1}$ as given by CKV) is related to the determination of the zero in the momentum scale whereas $\pm 0.004$ \AA$^{-1}$ gives the overall accuracy of the experiment\cite{Accuracy}, and the latter, in fact, includes the former. Further, in their attempt to observe the desired effect in our data by manipulation of the curves from the Figs. 3 e, f (Ref. \onlinecite{CPOD}), CKV make a series of factual errors. First of all, the room temperature curve was shown in Fig. 3 e (Ref. \onlinecite{CPOD}) only to demonstrate the application of our additional criterion (see below) and therefore was far from being central in our discussion. Nevertheless, we compare again our data with the data of Kaminski et al. in Fig.~\ref{Comparison} here, this time also in the same momentum range. Apparent qualitative and quantitative disagreement with the results presented in Fig. 3 g in Ref. \onlinecite{Kaminski} is clearly seen. The linear fit to the room temperature data points crosses the horizontal axis at 0.009 \AA$^{-1}$. This value, being equal to the shift of the extrema of the full-range curve shown in Fig. 3 e (Ref. \onlinecite{CPOD}) most likely originates from the residual misalignments at room temperature as was explained in the manuscript\cite{CPOD} on page 4. Moreover, it is the {\it room temperature curve} that appears to be displaced giving the approximately seven times smaller dichroism in the mirror plane (or nearly four times smaller shift in \AA$^{-1}$, not "about 1/2" as claimed by CKV) than the one observed in Ref. \onlinecite{Kaminski} in the pseudogap regime. Disregarding our explanation in terms of the misalignments one may still try to consider the relative shift of the point on the momentum axis for which the dichroism is zero as a measure of the "effect" which though sets in at room temperature. Needless to say that the error bars for this quantity, which is a difference of two other quantities determined in separate experiments, are calculated as a sum of the error bars corresponding to these two experiments. Careful reading of the manuscript\cite{CPOD} (page 5) reveals that the error bars of $\pm 0.005$ \AA$^{-1}$ are given for these particular cases. Thus the relative shift of the crossing point on the momentum axis of 0.01 \AA$^{-1}$ (see Fig.~\ref{Comparison}) is within the total error bars ($\pm 0.01$ \AA$^{-1}$). Therefore the statement that we have observed the "anticipated effect" is not correct.

When listing the reasons for non-observation of a similar effect in our experiments on page 5 of their preprint, CKV stated that the overall dichroism caused by the geometric effect is twice weaker than that observed earlier\cite{Kaminski} and related this to the smaller degree of the circular polarization, wrongly claiming that no information about this is given in our manuscript. Again, the choice of the data sets for the comparison is poor. The data shown in Fig. 3 e and f are taken in the second Brillouin zone. To compare the magnitude of the dichroic signal due to the geometric effect it is instructive to compare, e.g., the data from the overdoped samples shown in Fig. 2 d (Ref. \onlinecite{CPOD}) and Fig. 3 c (Ref. \onlinecite{Kaminski}), which are taken under comparable conditions. The quantitative agreement is remarkable, which is not surprising since the degree of the circular polarization used in our study is only slightly higher  (90$\%$ vs. 86$\%$). The precise location of the wanted information in our manuscript follows: page 1, last paragraph, first sentence.

There is no systematic shift also in the two data sets presented in Fig. 3 c in Ref. \onlinecite{CPOD} (Fig.~\ref{Key} a, b here). As already mentioned in our manuscript\cite{CPOD}, at zero momentum the linear fits cross the vertical axis at 0.16$\%$ and -0.41$\%$, respectively, both values being within the error bars of $\sim 0.5 \%$. Again, if one considers the relative shift, it is easy to see that the value of 0.16+0.41=0.57$\%$ is within the total error bars calculated as a sum of the error bars corresponding to the two key experiments.

As stated above and in accordance with the earlier experiments\cite{Kaminski}, predictions\cite{Varma_old} and interpretations\cite{Simon}, we believe that the key experiments should be carried out in the pseudogap state despite the effort of CKV to gradually "adjust" the criterion from the check of the reflection symmetry in the pseudogap state to the shift of the $\bf D(\bf k)$ curve with temperature in Ref. \onlinecite{CKV}. Naturally, auxilary experiments are needed to ensure the generality of the observations, for instance, like those we show in Ref. \onlinecite{CPOD} for the overdoped sample where the effect is not expected. One may also check whether the effect is zero for, e. g., a polycrystalline gold sample but the absence of higher temperature data in Fig. 3 d (Ref. \onlinecite{CPOD}) by no means depreciates the significance of the key experiment. We note, that such "zero-checks" become vitally important when a non-zero value of dichroism is detected in the mirror plane, just as in Ref. \onlinecite{Kaminski}. From this point of view our experiments can be considered as "zero-check" performed on 5x1 superstructure-free crystals\cite{Superstructure}.

Another concern of CKV is the intensity of the maximum of the full-range curve shown in Fig. 3 d (Ref. \onlinecite{CPOD}) as a function of temperature. It is not clear to us why such routinely observed variations of the dichroic signal away from the mirror plane with temperature (also seen in Fig. 3 c in Ref. \onlinecite{Kaminski}) "should be equal within the stated error bar of 0.05$\%$". We have never stated either this or the error bars of 0.05$\%$ for this relatively unimportant momentum region for the effect.

Contrary to what is stated in Ref. \onlinecite{CKV}, we neither consider the momentum coordinate of zero $\bf D_{N}$ equivalent to 0 nor give any error bars for the data shown in Fig. 4 b (Ref. \onlinecite{CPOD}). Instead, we explain that the alignment of the sample is extremely difficult in this particular situation (which obviously implies larger errors) and that there is no evidence for dramatic temperature dependent changes in the region of interest. 

\begin{figure}[t!]
\includegraphics[width=12cm]{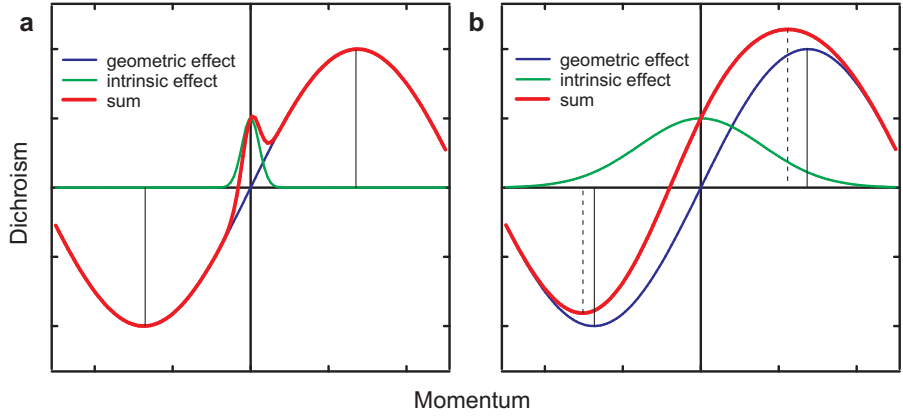}
\caption{\label{Lineshape} Sketch of the expected dichroism $\bf D_{N}(\bf k)$ in the presence of an even intrinsic effect maximal in the mirror plane. a) Narrow signal results in significant variations of the lineshape. b) Wider signal shifts the maximum and minimum of $\bf D_{N}(\bf k)$.}
\end{figure}

The lineshape analysis criterion developed by us can also be used as an alternative to the absolute intensity criterion described above. We have found that the $\bf D_{N}$ and {\bf D}-curves, when recorded in a significantly wider momentum range than in Ref. \onlinecite{Kaminski}, are nonmonotonic and in certain cases possess well defined extrema. In addtion, it turned out that these extrema are symmetrically located if the sample is properly aligned. The equations in our manuscript show that the lineshape of the $\bf D_{N}$-curve is essentially insensitive to the parameter $\alpha$, the ratio of the photon fluxes for plus and minus helicities, which is crucial for the absolute intensity method. The criterion works as follows. According to the expectations, the sought effect should be {\it maximal} in the ($\pi$, 0)-point (Ref. \onlinecite{Simon}, page 3). We illustrate two possible scenarios in Fig.~\ref{Lineshape} where the superposition of the geometric and intrinsic signals is seen to exhibit considerable variations in its lineshape. A narrow intrinsic effect would produce noticeable features near the mirror plane whereas the wider one would shift the extrema. Note, that to judge whether the effect is present or not, one does not require precise knowledge of the vertical scale. The idea of our approach is based on the trivial fact that the sum of an odd and even functions would inevitably break the reflection antisymmetry of the geometric dichroism. There is however one exception when our approach does not work. One may declare now that the sought signal should be constant as a function of {\bf k}. This seems to be suggested in Ref. \onlinecite{CKV}, to quote the paper  "... Only the value of the whole curve $\bf D(\bf k)$, including at f1 and f2 should move with respect to that without time-reversal violation ...". We want to notice that the momentum range in which we measure the dichroism is typically $\sim$ 0.3 \AA$^{-1}$ which is more than a third of the $\Gamma$-($\pi$, 0) distance. The assumption that the sought effect should be {\bf k}-independent on such a scale is in apparent contradiction with the theoretical expectations\cite{Simon}.

Exactly because of such possible "variations" in the description of the intrinsic dichroism we did not make any decisive conclusions based purely on this method in the manuscript, contrary to what is stated by CKV. We only used it partially and in some cases, checking momentum positions of the maxima and minima to monitor misalignments. In particular, the displacement of the room temperature curve heavily referenced by CKV from Fig.3 e (Ref.\onlinecite{CPOD}), has been shown to originate from a misalignment as both extrema were shifted in a characteristic way. Obviously enough, the statements of CKV about the "misconception" are not supported by the facts.

To summarize, we have presented the main results of our previous study which unambiguously proove the absence of the effect in Pb-Bi2212. We have also addressed all questions rised by CKV in their recent preprint. It turns out that {\it every} critical comment is either erroneous or irrelevant, thus being not able to question the validity of our results. 


\begin{thebibliography}{}
\bibitem{First_dichro} It was predicted that the ARPES "... at a specific point near the Fermi surface should have an electron yield which is different for right circularly polarized and left circularly polarized photons if the ground state has T breaking of the form shown in Fig. 2. Furthermore, the relative intensity should change in a systematic fashion with the momentum around the Fermi surface." For details see Ref.\onlinecite{Varma_old}.
\bibitem{Planes} Provided the vectors of incidence, emission and normal to the surface are coplanar.
\end{thebibliography}
\end{document}